\documentclass[aps,prl,twocolumn,showpacs,tightenlines,superscriptaddress]{revtex4}
\usepackage{epsfig,rotating}

\newcommand{\nuc}[2]{\hbox{$^{#1}$#2}}

\begin{document}
\title{Evolution of the $E(1/2^+_1)-E(3/2^+_1)$ energy spacing in
  odd-mass K, Cl and P isotopes for $N=20-28$  }

\author{A.~Gade}
   \affiliation{National Superconducting Cyclotron Laboratory,
     Michigan State University,
     East Lansing, MI 48824}
   \affiliation{Department of Physics and Astronomy,
     Michigan State University,
     East Lansing, MI 48824}
\author{B.\,A.~Brown}
   \affiliation{National Superconducting Cyclotron Laboratory,
     Michigan State University,
     East Lansing, MI 48824}
   \affiliation{Department of Physics and Astronomy,
     Michigan State University,
     East Lansing, MI 48824}
\author{D.~Bazin}
   \affiliation{National Superconducting Cyclotron Laboratory,
     Michigan State University,
     East Lansing, MI 48824}
\author{C.\,M.~Campbell}
   \affiliation{National Superconducting Cyclotron Laboratory,
     Michigan State University,
     East Lansing, MI 48824}
   \affiliation{Department of Physics and Astronomy,
     Michigan State University,
     East Lansing, MI 48824}
\author{J.\,A.~Church}
   \affiliation{National Superconducting Cyclotron Laboratory,
     Michigan State University,
     East Lansing, MI 48824}
   \affiliation{Department of Physics and Astronomy,
     Michigan State University,
     East Lansing, MI 48824}
\author{D.\,C.~Dinca}
   \affiliation{National Superconducting Cyclotron Laboratory,
     Michigan State University,
     East Lansing, MI 48824}
   \affiliation{Department of Physics and Astronomy,
     Michigan State University,
     East Lansing, MI 48824}
\author{J.~Enders}
   \affiliation{Institut f\"ur Kernphysik,
    Technische Universit\"at Darmstadt, Germany}
\author{T.~Glasmacher}
   \affiliation{National Superconducting Cyclotron Laboratory,
     Michigan State University,
     East Lansing, MI 48824}
   \affiliation{Department of Physics and Astronomy,
     Michigan State University,
     East Lansing, MI 48824}
\author{M.\ Horoi}
   \affiliation{Department of Physics, Central Michigan
     University, Mount Pleasant, MI 48859}
\author{Z.~Hu}
   \affiliation{National Superconducting Cyclotron Laboratory,
     Michigan State University,
     East Lansing, MI 48824}
\author{K.\,W.~Kemper}
   \affiliation{Department of Physics, Florida State University,
     Tallahassee, FL 32306}
\author{W.\,F.~Mueller}
   \affiliation{National Superconducting Cyclotron Laboratory,
     Michigan State University,
     East Lansing, MI 48824}
\author{T.~Otsuka}
    \affiliation{Department of Physics and Center for Nuclear Study,
     University of Tokyo, Hongo, Tokyo 113-0033, Japan}
    \affiliation{RIKEN, Hirosawa, Wako-shi, Saitama 351-0198, Japan}
\author{L.\,A.~Riley}
    \affiliation{Department of Physics and Astronomy, Ursinus College,
     Collegeville, PA 19426}
\author{B.\,T.~Roeder}
    \affiliation{Department of Physics, Florida State University,
     Tallahassee, FL 32306}
\author{T.\ Suzuki}
    \affiliation{Department of Physics, Nihon University, Sakurajosui,
     Setagaya-ku, Tokyo, 156-8550, Japan}
\author{J.\,R.~Terry}
   \affiliation{National Superconducting Cyclotron Laboratory,
     Michigan State University,
     East Lansing, MI 48824}
   \affiliation{Department of Physics and Astronomy,
     Michigan State University,
     East Lansing, MI 48824}
\author{K.\,L.~Yurkewicz}
   \affiliation{National Superconducting Cyclotron Laboratory,
     Michigan State University,
     East Lansing, MI 48824}
   \affiliation{Department of Physics and Astronomy,
     Michigan State University,
     East Lansing, MI 48824}
\author{H.~Zwahlen}
   \affiliation{National Superconducting Cyclotron Laboratory,
     Michigan State University,
     East Lansing, MI 48824}
   \affiliation{Department of Physics and Astronomy,
     Michigan State University,
     East Lansing, MI 48824}

\date{\today}

\begin{abstract}
The energy of the first excited state in the neutron-rich $N=28$
nucleus \nuc{45}{Cl} has been established via in-beam $\gamma$-ray
spectroscopy following proton removal. This energy value completes the
systematics of the $E(1/2^+_1)-E(3/2^+_1)$ level spacing in odd-mass
K, Cl and P isotopes for $N=20-28$. The results are discussed in the
framework of shell-model calculations in the $sd$-$fp$ model space.
The contribution of the central, spin-orbit and tensor components is
discussed from a calculation based on a proton single-hole spectrum
from $G$-matrix and $\pi + \rho$ meson exchange potentials. A
composite model for the proton $0d_{3/2}-1s_{1/2}$ single-particle
energy shift is presented.
\end{abstract}

\maketitle
Neutron-rich nuclei in the neighborhood of \nuc{44}{S} have attracted
much attention in recent years. The question whether the high degree
of collectivity observed for \nuc{42,44}{S}~\cite{Sch96,Gla97} is due
to a breakdown of
the $N=28$ neutron-magic number or the collapse of the $Z=16$ proton
sub-shell gap at neutron number 28 is much discussed in the
literature~\cite{Wer94,Ret97,Cot98,Cot00,Soh02,Sor04}. The vanishing
of the $Z=16$ subshell closure was inferred from the near-degeneracy
of the proton $s_{1/2}$ and $d_{3/2}$ orbitals in the chain of K
isotopes as $N=28$ is approached~\cite{Ret97,Cot98,ots05}.

Retamosa {\it et al.}~\cite{Ret97} present an unrestricted shell-model
calculation in a valence space including the $sd$ shell for protons
and the $pf$ shell for neutrons. The evolution of the
$E(1/2^+_1)-E(3/2^+_1)$ level spacing  in the K isotopes was used to
phenomenologically modify the cross-shell interaction. The authors
predict the evolution of the $E(1/2^+_1)-E(3/2^+_1)$ energy difference
in the $Z=17$ and $Z=15$ isotopic chains as neutrons fill the
$f_{7/2}$ orbit. At that time, the $E(1/2^+_1)-E(3/2^+_1)$ energy
splitting was neither known in any of the P isotopes with $20 \leq N
\leq 28$ nor in the Cl isotopes above $N=22$. In the present paper, we
complete the systematics of the experimental $1/2^+_1 - 3/2^+_1$ level
spacings in the Cl chain. The contributions of the
central, spin-orbit and tensor components of the NN interaction to the
evolution of the energy splitting are analyzed to elucidate the
microscopic effects driving the changes in single-particle structure.
For this, single proton-hole spectra are discussed in the framework of
$G$-matrix and $\pi + \rho$ meson exchange potentials.

\begin{figure}[h]
        \epsfxsize 7.8cm
        \epsfbox{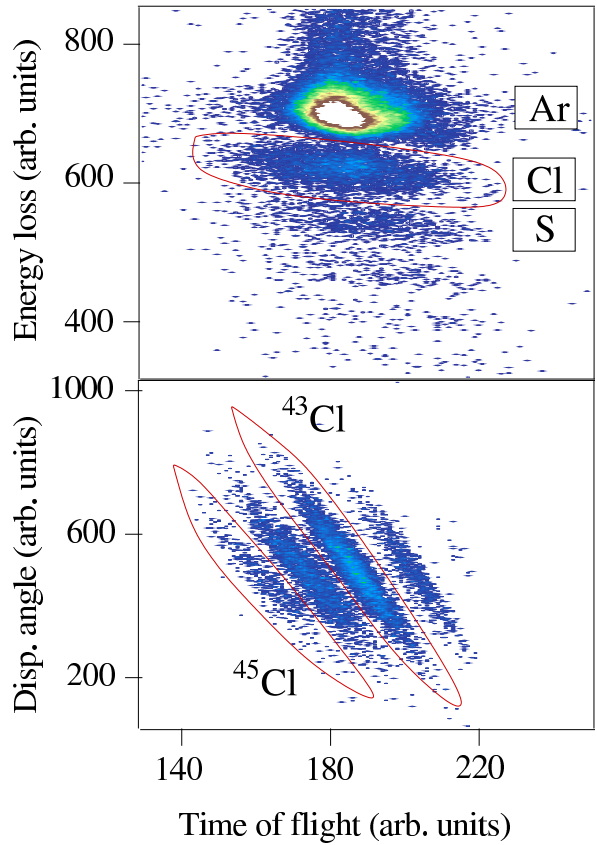}
\caption{\label{fig:pid} (Color online) Particle identification in the
  S800 focal 
  plane. The upper panel shows the energy loss measured in the ion
  chamber vs. time of flight taken between two scintillators. The
  lower panel shows for Cl isotopes the dispersive angle in the focal
  plane measured with the position sensitive CRDCs vs. the time of
  flight.} 
\end{figure}

The experiment was performed at the Coupled Cyclotron Facility of the
National Superconducting Cyclotron Laboratory at Michigan State
University. The 76.4 MeV/nucleon \nuc{46}{Ar} secondary beam was
produced via projectile fragmentation of a 110~MeV/nucleon
\nuc{48}{Ca} primary beam on a 376~mg/cm$^2$ \nuc{9}{Be} target
located at the mid-target position of the A1900 fragment
separator~\cite{a1900}. The separator was operated with 0.5\% momentum
acceptance and a beam purity of about 99\% was achieved. The
\nuc{46}{Ar} secondary beam was incident on a 191~mg/cm$^2$
polypropylene [(C$_3$H$_6)_n$] target at the target position of the
S800 spectrograph~\cite{s800}. The reaction products were identified
event-by-event with the spectrograph's focal-plane detector
system~\cite{Yur99} in conjunction with time-of-flight information
obtained from scintillators in the beam line. Figure \ref{fig:pid}
shows the particle identification for the Cl isotopes studied in this
experiment. The energy-loss information from the S800 ionization chamber
provides $Z$ identification (upper panel). For given isotope, the
correlation between the dispersive angle in the S800 focal plane and
the time-of-flight information resolves $A$ (lower panel). 

The magnetic rigidity of
the spectrograph was centered on the inelastic scattering of
\nuc{46}{Ar} off the polypropylene target (see Ref.~\cite{Ril05}).
However, the momentum acceptance of the S800 spectrograph was large
enough to allow a fraction of the one-proton knockout residues
\nuc{45}{Cl} and the multi-nucleon removal residues \nuc{43}{Cl} to
enter the focal plane as well. Only the tail of the \nuc{45}{Cl}
momentum distribution was within the acceptance, confining the present
study to in-beam $\gamma$-ray spectroscopy.   

The target was surrounded by SeGA, an array of 32-fold segmented,
high-purity Ge detectors~\cite{sega} arranged in two rings with angles
of 90$^{\circ}$ and 37$^{\circ}$ with respect to the beam axis,
respectively. Fifteen of the 18 SeGA detectors were used for the
present experiment. The high degree of segmentation is necessary to
Doppler reconstruct the $\gamma$ rays emitted by the
reaction residues in flight.

\begin{figure}[h]
        \epsfxsize 8.2cm
        \epsfbox{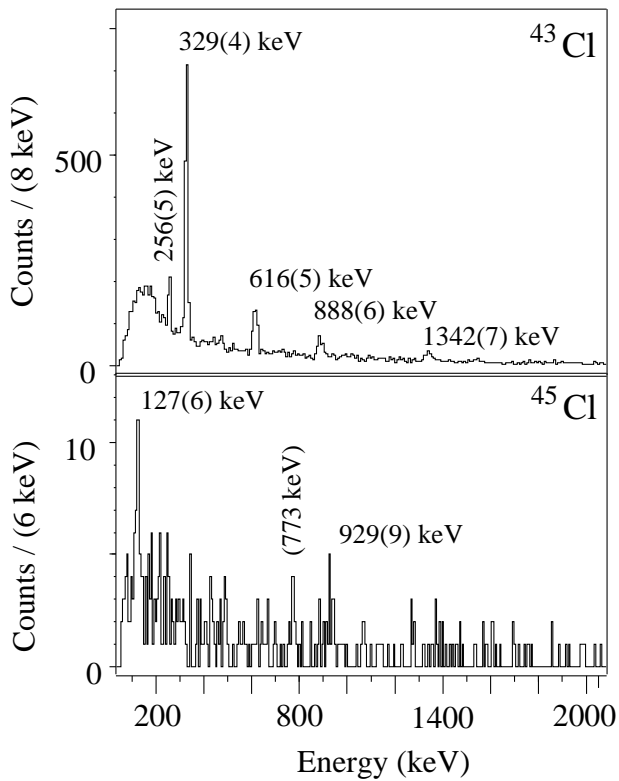}
\caption{\label{fig:spectra} Event-by-event Doppler-reconstructed
  $\gamma$-ray spectra in coincidence 
with \nuc{43}{Cl} and \nuc{45}{Cl} nucleon-removal residues produced
from an \nuc{46}{Ar} secondary beam impinging on a polypropylene target.}
\end{figure}

The upper panel of Fig.~\ref{fig:spectra} shows the $\gamma$-ray
spectrum detected in coincidence with \nuc{43}{Cl} produced by
multi-nucleon removal from the \nuc{46}{Ar} secondary beam incident on
the polypropylene target. The $\gamma$ rays at 329(4)~keV, 616(5)~keV,
888(6)~keV and 1342(7)~keV observed in \nuc{43}{Cl} are in agreement
with transitions reported in~\cite{Sor04} from \nuc{48}{Ca}
fragmentation at 60~MeV/nucleon. In addition, we see a $\gamma$-ray
transition at 
256(5)~keV that would have been difficult to be detected
by~\cite{Sor04} due to their fairly high detection threshold for
$\gamma$-ray energies (see Fig. 4 of Ref.~\cite{Sor04} showing the
$\gamma$-ray spectra of \nuc{43}{Cl} and \nuc{45}{Cl} detected with
segmented Ge detectors of the Clover type following fragmentation of
\nuc{48}{Ca}).  The 
1509(10)~keV $\gamma$-ray peak observed by Sorlin {\it et
al.}~\cite{Sor04} might correspond to the decay of a state that is
populated in the fragmentation of \nuc{48}{Ca} but unaccessible from
nucleon removal of \nuc{46}{Ar} projectiles.

\begin{figure}[h]
        \epsfxsize 8.4cm
        \epsfbox{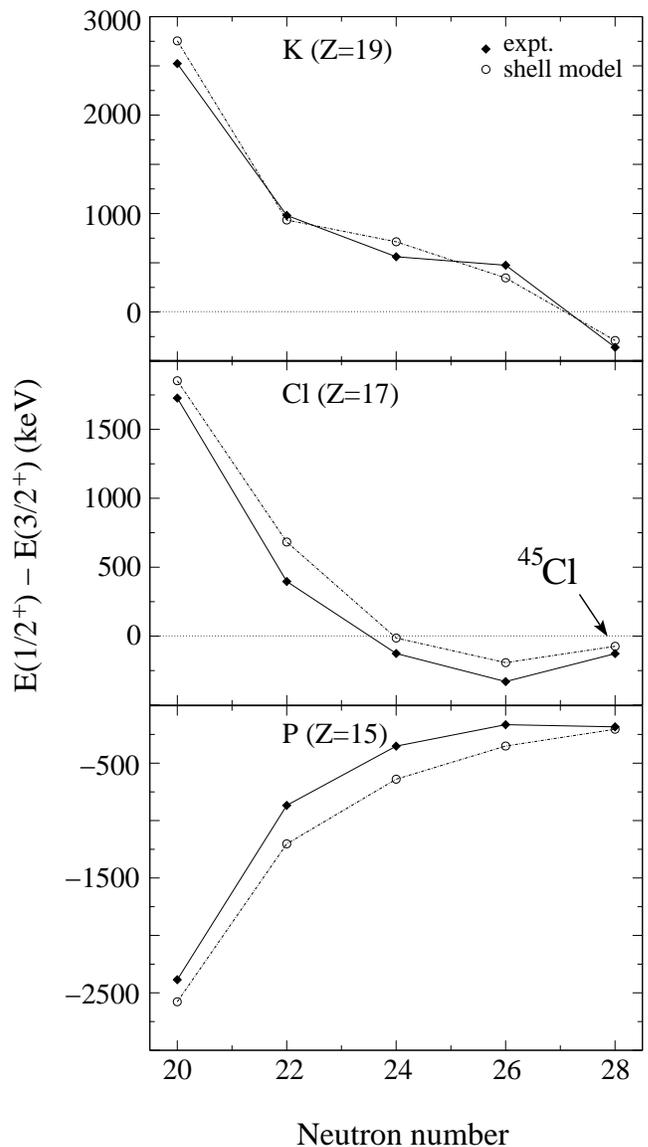}
\caption{\label{fig:comp_sm} Comparison of the experimental
  $\Delta_{13}=E(1/2^+_1)-E(3/2^+_1)$ energy splitting to shell-model
  calculations
  using the Nowacki effective interaction~\cite{Num01}. The ordering of the
  $1/2^+$ and $3/2^+$ levels in \nuc{41}{P}~\cite{Cam05} and
  \nuc{45}{Cl} has not
  been determined by experiment and is assigned by comparison with
  calculations. The value for \nuc{43}{P} stems from~\cite{Fri05},
  others from~\cite{Sor04,kramer}.}
\end{figure}

The lower panel of Fig.~\ref{fig:spectra} displays the $\gamma$ rays
in coincidence with the \nuc{45}{Cl} one-proton knockout residues. The
929(9)~keV $\gamma$-ray corresponds to the transition previously
observed in intermediate-energy Coulomb excitation employing a NaI
array for $\gamma$-ray detection~\cite{Ibb99}.  The
existence of a peak at 773 keV is less clear. The dominant
$\gamma$-ray transition in our spectrum is found at 127(6)~keV and is
attributed to a transition between the $3/2^+_1$ and $1/2^+_1$ states.
Shell-model calculations predict the ground state of \nuc{45}{Cl} to
be $1/2^+$ with the first excited $3/2^+_1$ state at 74~keV. Our
experimental result is in agreement with this expected energy
splitting between the $1/2^+_1$ and $3/2^+_1$ states and completes the
systematics of $\Delta_{13}$ in the chain of Cl isotopes for $20 \leq
N \leq 28$. This 127(6)~keV $\gamma$ ray could not be observed by
Sorlin {\it et al.} due to their high detection threshold (see Fig. 4
of~\cite{Sor04}). The evolution of the energy difference
$E(1/2^+_1)-E(3/2^+_1)$ in the chains of K, Cl and P isotopes for
neutron numbers from $N=20-28$ is shown in Fig.~\ref{fig:comp_sm} and
compared to shell-model calculations using the Nowacki
interaction~\cite{Num01}. In our calculation, the protons are
confined to the $sd$ shell, $sd$-shell neutrons are in the
closed-shell configuration $\nu(sd)^{12}$ and the remaining neutrons
occupy the $fp$ shell. In this space, \nuc{48}{Ca} has the configuration
$\pi(sd)^{12}~\nu(sd)^{12}~\nu(pf)^8$. In Table~\ref{tab:occ} we give
the $sd$ shell occupation of the discussed $1/2^+_1$ and $3/2^+_1$
states in the Cl isotopes.

\begin{table}[h]
\begin{center}
 \vspace{0.5cm}
\caption{Proton shell-model occupancies $n$ for the lowest-lying  $1/2^+$ and
  $3/2^+$ states in chain of Cl isotopes. The rather high and constant
  $d_{5/2}$ occupancy is an indication of the spherical nature of
  these nuclei.\label{tab:occ}}
\begin{ruledtabular}
\begin{tabular}{ccccc}
   & $J^{\pi}$ & $n(d_{5/2})$ & $n(d_{3/2})$ & $n(s_{1/2})$\\
\hline
\nuc{37}{Cl} & $1/2^+$ & 5.91 & 2.07 & 1.02 \\
             & $3/2^+$ & 5.93 & 1.12 & 1.95 \\
\nuc{39}{Cl} & $1/2^+$ & 5.89 & 1.89 & 1.22 \\
             & $3/2^+$ & 5.89 & 1.31 & 1.80 \\
\nuc{41}{Cl} & $1/2^+$ & 5.86 & 1.90 & 1.24 \\
             & $3/2^+$ & 5.86 & 1.54 & 1.60 \\
\nuc{43}{Cl} & $1/2^+$ & 5.83 & 1.96 & 1.21 \\
             & $3/2^+$ & 5.85 & 1.94 & 1.21 \\
\nuc{45}{Cl} & $1/2^+$ & 5.85 & 2.22 & 0.93 \\
             & $3/2^+$ & 5.91 & 2.36 & 0.73 \\
\end{tabular}
\end{ruledtabular}
\end{center}
\end{table}

 We first
analyze the difference between the $d_{3/2}$ and $s_{1/2}$
proton-removal energies from Ca to K, $\Delta_{13}$, in terms of its
dependence on the interaction components. The experimental values are
given in Table~\ref{tab:delta13}. The energies of the lowest $1/2^+$
and $3/2^+$ states in \nuc{47}{K} give  
$\Delta_{13}=-0.36$~MeV.
The centroid energy of the $s_{1/2}$ and $d_{3/2}$ strength from the
\nuc{48}{Ca}$(e,e'p)$\nuc{47}{K} 
experiment of reference~\cite{kramer} is $\Delta_{13} = -0.29$~MeV. Previous
comparisons (references \cite{Cot00} and \cite{ots05}) have used a
value of $\Delta_{13} =+0.29$~MeV based on the older 
\nuc{48}{Ca}$(d,\nuc{3}{He})$ experiment of \cite{doll}. However, the
$\ell=2$ strength 
reported in Ref. \cite{doll} could be either attributed to $d_{3/2}$ or $d_{5/2}$ and it was
simply assumed that all states except the ground state were of spin
and parity $5/2^+$.
In reference \cite{kramer} the value of -0.29 is based on new
$J^{\pi}$ assignments  
given in Table I of that paper, but it is not clear to 
us if these assignments are firm. In the shell-model
calculations the lowest energy-spacing is
-0.31~MeV compared to the centroid energy spacing of -0.17~MeV.

In order to have a microscopic interpretation of the results we have
calculated the single-hole spectrum for protons from a $G$-matrix
potential \cite{h7b} based on the Paris NN potential. The results are
given in Table~\ref{tab:delta13} broken down into the contributions of
the central, spin-orbit and tensor components of the interaction.
It has been shown that the monopole part of the $G$ matrix is not
so reliable \cite{McG70,Pov81,Bro88}; therefore, it is of interest how
the individual
contributions compare to other calculations. The importance of the
spin-isospin part of the NN
interaction has been pointed out in~\cite{otsuka} for the changes of
the shell structure across the nuclear chart.
It is worth mentioning that
the monopole part of the tensor force has been shown in \cite{ots05}
to change the shell structure in a unique and robust way
across the nuclear chart.
Table~\ref{tab:delta13} shows the effect of the tensor part of the present $G$-matrix
calculation and the tensor contribution as derived from the one-$\pi$
and one-$\rho$ meson
exchange tensor potential similar to
\cite{ots05} for $A=40$. One notices that the two tensor results are
remarkably
close to each other.  This is in fact an example of the universality
of the tensor monopole effect from its longer range part as pointed
out in \cite{ots05}.


\begin{table}[h]
\caption{Splitting between the $d_{3/2}$ and $s_{1/2}$ proton hole
energies $\Delta_{13}$ in units of MeV. The result for the $G$ matrix
calculation is
decomposed into the central, spin-orbit and tensor contribution.}
\begin{ruledtabular}
\begin{tabular}{lccc}
 $\Delta_{13}$  &  \nuc{39}{K} & \nuc{47}{K} & \nuc{39}{K} - \nuc{47}{K}\\
    (MeV)       &              &             &               \\
\hline
``expt.''\footnote[1]{$E(1/2^+_1)-E(3/2^+_1)$}            & 2.52  &
    -0.36 & 2.88  \\
shell model~\footnote[3]{with the Nowacki effective interaction~\cite{Num01}} & 2.75  & -0.31 & 3.06 \\
\hline
$G$ matrix total       & 3.66  & -0.73 & 4.39 \\
(central)            & 0.98  & -1.28 & 2.26 \\
(spin-orbit)         & 2.68  &  2.10 & 0.58 \\
(tensor)             & 0.00  & -1.55 & 1.55 \\
\hline
$\pi + \rho$ tensor~\cite{ots05}        & 0.00  & -1.67 & 1.67
\label{tab:delta13}
\end{tabular}
\end{ruledtabular}
\end{table}

The tensor part can be further examined by the $d_{5/2}-d_{3/2}$
spin-orbit splitting, $\Delta_{53}$, as shown in
Table~\ref{tab:delta53}. The experimental energy is the energy
centroid of the $d_{5/2}$ hole strength observed in
\nuc{40}{Ca}~\cite{doll} and \nuc{48}{Ca}~\cite{kramer}. The Nowacki
interaction results are again based on the ($f_{7/2}$)$^{8}$ neutron
configuration. (The centroid energy from the full $pf$-shell model
space is 5.76~MeV.) One observes a decrease in the experimental
spin-orbit interaction that, when compared to the $G$-matrix
calculation, is mainly attributed to the tensor interaction,
consistent with Ref. \cite{ots05}. In fact, Table~\ref{tab:delta53}
indicates that the result of the one-$\pi$ and one-$\rho$ meson
exchange tensor potential is in very good agreement with the
experiment.

The absolute
spin-orbit interaction obtained with the $G$-matrix interaction
 in \nuc{40}{Ca}
amounts  only for about 60\% of the experimental value (first column
of Table~\ref{tab:delta53}). It has been shown that the spin-orbit
splitting can be 
reproduced by a microscopic calculation based on the UMOA method from
the bare NN interaction for $^{16}$O \cite{fujii}. In this
calculation, more complex components are included but their effects
are renormalized in the conventional shell-model picture. The
three-body interaction has been shown also to contribute to the
spin-orbit splitting in light nuclei \cite{navratil}. Thus, contrary
to the tensor force, the relation between the spin-orbit force and the
splitting remains to be clarified.

\begin{table}[h]
\caption{Splitting between the $d_{5/2}$ and $d_{3/2}$ proton hole
energies $\Delta_{53}$ in units of MeV. The result for the $G$ matrix is
decomposed into the central, spin-orbit and tensor contribution.}
\begin{ruledtabular}
\begin{tabular}{lccc}
 $\Delta_{53}$  &  \nuc{39}{K} & \nuc{47}{K} & \nuc{39}{K} - \nuc{47}{K}\\
    (MeV)       &              &             &               \\
\hline
``expt.''\footnote[3]{energy centroids from \cite{doll,kramer}}  & 7.5   & 4.8   & 2.7  \\
shell model~\footnote[2]{with the Nowacki effective interaction~\cite{Num01}} & 7.4  & 5.92 & 1.48 \\
\hline
$G$ matrix total       & 3.94  & 0.84 & 3.10 \\
(central)            & 0.00  & -0.32 & 0.32 \\
(spin-orbit)         & 3.94  &  3.86 & 0.08 \\
(tensor)             & 0.00  & -2.70 & 2.70 \\
\hline
$\pi + \rho$ tensor~\cite{ots05}        & 0.00  & -2.78 & 2.78
\label{tab:delta53}
\end{tabular}
\end{ruledtabular}
\end{table}

The Skyrme Hartree-Fock (HF) method can also be used to calculate the central
interaction
contribution to $\Delta_{13}$ (this is done by calculating the single-particle
spectrum with the Skyrme spin-orbit strength set to zero). The values
from the Skyrme SKX~\cite{Bro98} HF calculation are given in the
second row of Table~\ref{tab:skx}.
The Skyrme results can differ from the $G$-matrix values due to finite well and
density-dependent (or implicit effective three-body) effects.
However, this HF does not include the
tensor contribution.

Taking all of these into account we might make a
composite model of the single-particle shifts based on HF for central,
$G$-matrix for spin-orbit (Table~\ref{tab:delta13})
and $\pi + \rho$ for tensor contributions. The
results as given in Table~\ref{tab:skx}  are in reasonable agreement
with experiment when the spin-orbit part from the $G$ matrix is scaled
by a factor of 1.9 as obtained from Table~\ref{tab:delta53}.
The need for rescaling the spin-orbit part is mainly
due to monopole effects only inaccurately taken into account.
We note that the monopole effect from the central potential
differs considerably between the $G$-matrix and SKX interactions,
which implies intrinsic theoretical difficulties.
The relative importance of the central and spin-orbit potentials
cannot be clarified in the present study, however, their combined
effect seems to be about half of the tensor monopole effect
for $\Delta_{13}$, while negligible for $\Delta_{53}$.
A more precise evaluation of their magnitude and interplay remains an
intriguing problem. The $1/2^+$ proton (Nilsson) state, which is the
highest $K=1/2^+$ of $sd$-shell origin, can be pushed up due to
deformation. This would result in a lower energy of the $1/2^+$ level in
the observed spectrum of the actual nucleus. This could occur more
easily as $d_{3/2}$ and $s_{1/2}$ come closer in energy (i.e.,
stronger mixing). Thus, in this case, the ``experimental''
$\Delta_{13}$ would appear larger than the pure single-particle
effect. This point should be taken into consideration more precisely
in the future.


\begin{table}[h]
\caption{Splitting between the $d_{3/2}$ and $s_{1/2}$ proton hole
energies $\Delta_{13}$ in units of MeV compared to a composite model
of the single-particle shifts. The central part is obtained from the SKX
Skyrme HF calculation, the spin-orbit part is taken from the
$G$-matrix approach of
Table~\ref{tab:delta13} and the tensor contribution is based on the $\pi
+ \rho$ potential~\cite{ots05}. The spin-orbit contribution is scaled
by a factor
of 1.9 obtained from Table~\ref{tab:delta53}. }
\begin{ruledtabular}
\begin{tabular}{lccc}
 $\Delta_{13}$  &  \nuc{39}{K} & \nuc{47}{K} & \nuc{39}{K} - \nuc{47}{K}\\
    (MeV)       &              &             &                      \\
\hline
``expt.''\footnote[1]{$E(1/2^+_1)-E(3/2^+_1)$} & 2.52  & -0.36 & 2.88  \\
\hline
total                                        &  3.00  & -0.43 & 3.43 \\
(Skyrme central)                             & -2.09  & -2.75 & 0.66 \\
(1.9 $\times$ $G$-matrix spin-orbit)         &  5.09  &  3.99 & 1.10 \\
($\pi + \rho$ tensor)                        & 0.00   & -1.67 & 1.67
\label{tab:skx}
\end{tabular}
\end{ruledtabular}
\end{table}

In summary, we report on the first determination of the
$|E(1/2^+_1)-E(3/2^+_1)|=127(6)$~keV energy splitting in the $N=28$
nucleus \nuc{45}{Cl} observed following the one-proton removal from a
\nuc{46}{Ar} secondary beam upon collision with a polypropylene
target. The evolution of the energy splitting is compared to
shell-model calculations in the $sd$-$fp$ model space. Its dependence
on the interaction components, central, spin-orbit and tensor, is
discussed for the chain of K isotopes from calculations based on the
$G$ matrix and $\pi + \rho$ tensor potential. A similar analysis is
performed for the splitting between the $d_{5/2}$ and the $d_{3/2}$
orbit where the experimental determination of the location of the
$d_{5/2}$ single-particle strength in P and Cl has to remain a
challenge for future experiments. The tensor monopole effect is seen
as almost the sole source of the change of the $d_{5/2} - d_{3/2}$
spin-orbit splitting, while the central potential shows a certain
effect for the change of the $s_{1/2} -d_{3/2}$ spin-orbit splitting.
The change of the $1/2^+ - 3/2^+$ splitting contains more
uncertainties in relation to single-particle properties and needs
further studies. In this respect, the present experiment can be a
first step towards a more comprehensive understanding of this region.


\begin{acknowledgments}
Valuable discussions with F.\ Nowacki are
acknowledged. We thank A.\ Stolz, T.\ Ginter, M.\ Steiner and the NSCL
cyclotron operations group for providing the high-quality secondary
and primary beams. This work was supported by the National Science
Foundation under Grants No. PHY-0110253 and PHY-0244453. This work was
supported in part by a Grant-in-Aid for Specially Promoted Research
(No. 13002001) from the MEXT.
\end{acknowledgments}

\end{document}